# Packing grains by thermally cycling

One of the oldest and most intriguing problems in the handling of materials is how a collection of solid grains packs together.[1] While granular packing is normally determined by how grains are poured or shaken, we find that a systematic and controllable increase in packing is induced by simply raising and lowering the temperature, e.g., without the input of mechanical energy. The results demonstrate that thermal processing provides a largely unexplored mechanism of grain dynamics, as well as an important practical consideration in the handling and storage of granular materials.

The packing fraction of a granular material is defined as the fraction of sample volume filled by grains rather than by empty space. The packing fraction can typically vary between 57% and 64% for randomly arranged spherical grains and even more widely for other grain shapes. Packing is often made more dense through vibration or tapping, which induce small rearrangements and thus allow the grains to settle.[2,3,4]

Upon heating a granular material, the grains and their container both undergo thermal expansion. This can lead to settling due to the metastable nature of disordered grain configurations (especially if the grains and their container are made of different materials), and such settling should not be reversible upon cooling to ambient temperature. Previous studies have shown that temperature changes affect silos in industrial settings[5,6] and the stress state of a granular pile,[7,8] but the grain dynamics induced by thermal cycling has not been systematically explored.

We examined the change of packing fraction (ΔPF) due to both single thermal cycles and repeated cycles to the same temperature for glass spheres contained in vertical plastic cylinders (experimental details in supplementary material). As shown in Figure 1a, there is a clear increase in packing even for a single cycle to 10 degrees above ambient room temperature. The results were not affected by the height to which the cylinders were filled (to within ± 10%), the heating rate, or the time spent at the cycle temperature after thermal equilibrium is reached, and they changed only slightly (< 20%) as the cylinder diameter was changed by an order of magnitude (Figure 1b). This last result is physically sensible, since the expansion of both the grains and the container scales with the size of the sample. The packing fraction continues to increase under multiple successive thermal cycles (Figure 1c). The increase can be described by a double exponential relaxation consistent with a combination of large-scale and small-scale rearrangements (fits to the data are described in supplementary material).[9] The range of data does not allow us to definitively determine that this is the correct physical model, but the "time constants" of the two relaxations do increase with decreasing temperature – consistent with the smaller thermal expansion.

The primary cause of our observed ΔPF is most probably the difference between the thermal expansions of the container and the grains. Confirming this expectation, we observed similar changes in packing for plastic spheres contained in glass cylinders (where the grains expand more than the container), and smaller changes in packing for glass spheres contained in glass cylinders. Our data suggest the existence of a large manifold of possible thermal effects in granular media, analogous to the effects of vibration. Indeed, a geophysical form of granular segregation, "stone heave", has been associated with thermal effects,[10] and thermal effects are manifested in outdoor storage silos in which the grains become successively more packed with each diurnal cycle – potentially leading to catastrophic failure.[6] Perhaps most importantly, our data demonstrate an almost adiabatic alternative to mechanical agitation through which grain configurations can be altered, providing a new mechanism with which the very old subject of grain packing can be probed.

K. Chen, J. Cole, C. Conger, J. Draskovic, M. Lohr, K. Klein, T. Scheidemantel, and P. Schiffer

*Department of Physics and Materials Research Institute, Pennsylvania State University, University Park, PA 16802 USA*

e-mail: schiffer@phys.psu.edu

**Competing Interests statement**  The authors declare that they have no competing financial interests.

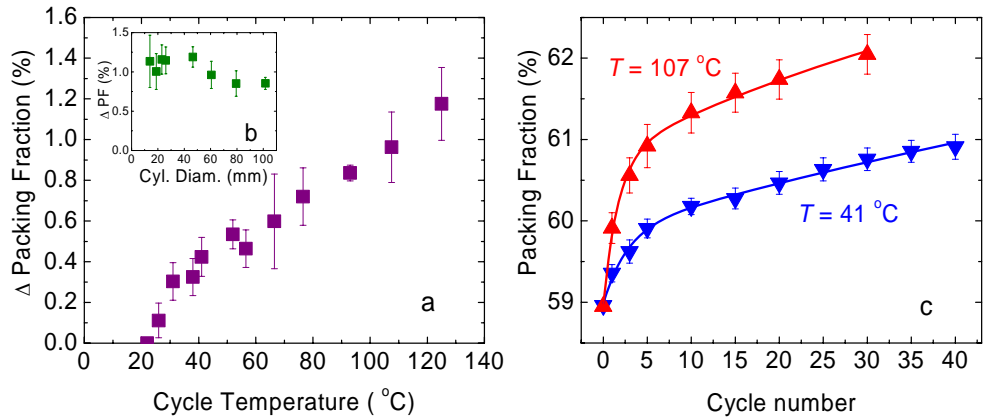

Figure 1. **Change of packing fraction with thermal cycling for glass spheres in a plastic cylinder**. **(a).** Change of packing fraction as a function of cycle temperature from room temperature for a single cycle (cylinder diameter 60.6 mm). **(b)**. Change of packing fraction as a function of cylinder diameter for a single cycle (cycle temperature of 107 ± 2 °C). **(c).** Evolution of packing fraction after multiple thermal cycles with cycle temperatures of 107 ± 2 °C (red up-triangle) and 41 ± 1 °C (blue down-triangle). Lines are fits to the data described in the text. The error bars in each case represent the standard deviation determined from several repeated measurements (typically twelve or more for (a) and (b) and six for (c)), and experimental details are given in supplementary information.

# Supplementary Information

## Supplementary methods

We prepared samples of 0.52 ± 0.06 mm diameter spherical soda lime glass beads (linear thermal expansion of $9\times10^{-6}$ K$^{-1}$) contained in graduated cylinders made of hard plastic (polymethylpentene, linear thermal expansion of $1.17\times10^{-4}$ K$^{-1}$). We started with initial packing fractions of 58.9 ± 0.1%, which is typical of grains poured into a container and not compacted. We then measured the increase in packing fraction (ΔPF) after samples were heated in air to a designated temperature, allowed to equilibrate at that temperature, and then allowed to cool to room temperature (22 ± 1°C). Within our temperature range, we expect the grains to expand approximately 0.48 μm (0.09% of particle diameter) at 125 °C, and except for the highest cycle temperatures, the cylinder expansion was considerably less than one bead diameter. We also performed similar experiments with plastic beads (polystyrene, linear thermal expansion of $7.3\times10^{-5}$ K$^{-1}$, diameter d = 0.98 ± 0.09 mm) in glass cylinders (borosilicate glass, linear thermal expansion of $3\times10^{-6}$ K$^{-1}$, diameter $D$ = 25 and 85 mm) and with soda lime glass beads in borosilicate glass cylinders.

Experimental checks were performed to ensure that there was no distortion of the container at the end of the cycling, that there was no vibration-induced packing, and that the results did not depend on the heating rate or time spent at the cycle temperature, as long as thermal equilibrium is reached at the cycle temperature.

A general two-mechanism model was proposed for density relaxation in shaken granular media by Barker and Mehta[1] in the form of

$$y = y_0 - A_1 e^{-\frac{x}{\tau_1}} - A_2 e^{-\frac{x}{\tau_2}}$$

where $\tau_1 < \tau_2$, $y_0$ is the maximum packing fraction, and $x$ is the number of shaking cycles. The shorter relaxation time $\tau_1$ corresponds to individual particle relaxation and the longer relaxation time $\tau_2$ corresponds to the relaxation of granular blocks. The magnitudes of these relaxations are characterized by $A_1$ and $A_2$ respectively. The data of repeated thermal cycles (Figure 1c) are fitted to this double-relaxation model with $y_0$ fixed to 64 (%), the maximum random packing fraction. Fitted parameters are listed below.

For cycle temperature of 107 °C,

$A_1 = 1.81 \pm 0.12$     $\tau_1 = 1.74 \pm 0.26$

$A_2 = 3.21 \pm 0.11$     $\tau_2 = 57.96 \pm 6.54$

reduced chi-square $\chi^2/DoF = 1.24$

For cycle temperature of 41 °C,

$A_1 = 0.90 \pm 0.07$     $\tau_1 = 2.72 \pm 0.48$

$A_2 = 4.11 \pm 0.06$     $\tau_2 = 131.79 \pm 10.18$

reduced chi-square $\chi^2/DoF = 1.11$

The reduced chi-square is defined as $\frac{1}{n-p} \sum_i \frac{(y_i - f(x_i))^2}{\sigma_i^2}$, where the $n$ is the number of experimental data points, $p$ is the number of parameters in the model, $y_i$ is the experimental packing fraction, $f(x_i)$ is the packing fraction predicted by the model and $\sigma_i$ is the standard error of experimental data.

Good agreement between experimental data and theoretical model is observed. For each cycle temperature two different relaxation times can be clearly identified. One is much smaller than the other. It is not surprising that both time constants increase when the cycle temperature is lowered, since more cycles are needed to achieve the same change of packing fraction at a lower temperature. Individual particle relaxation becomes more and more important as the cycle temperature is increased. This also agrees with the model proposed by Barker and Mehta.[1]

---

[1] Barker, G. C. & Mehta, A. Transient phenomena, self diffusion, and orientational effects in vibrated powders. *Phys. Rev. E* **47**, 184-188 (1993).